\documentclass[11pt,preprint,showpacs,preprintnumbers,amsmath,amssymb]{revtex4}

\usepackage{exscale}
\usepackage{relsize}
\usepackage{graphicx}
\usepackage{dcolumn}
\usepackage{bm}
\usepackage{multirow}
\usepackage{CJK}
\usepackage{diagbox}
\usepackage{subfigure}

\begin{document}

\title{Studying the localized $CP$ violation and the branching fraction of the $\bar{B}^0\rightarrow K^-\pi^+\pi^+\pi^-$ decay}

\author{Jing-Juan Qi \footnote{e-mail: jjqi@mail.bnu.edu.cn}}
\affiliation{\small{Junior College, Zhejiang Wanli University, Zhejiang 315101, China}}

\author{Zhen-Yang Wang \footnote{Corresponding author, e-mail: wangzhenyang@mail.nbu.edu.cn}}
\affiliation{\small{Physics Department, Ningbo University, Zhejiang 315211, China}}

\author{Jing Xu}
\affiliation{\small{Department of Physics, Yantai University, Yantai 264005, China}}

\author{Xin-Heng Guo \footnote{Corresponding author, e-mail: xhguo@bnu.edu.cn}}
\affiliation{\small{College of Nuclear Science and Technology, Beijing Normal University, Beijing 100875, China}}

\date{\today\\}

\begin{abstract}
In this work, we study the localized $CP$ violation and the branching fraction of the four-body decay $\bar{B}^0\rightarrow K^-\pi^+\pi^-\pi^+$ by employing a quasi-two-body QCD factorization approach. Considering the interference of $\bar{B}^0\rightarrow \bar{K}_0^*(700)\rho^0(770)\rightarrow K^-\pi^+\pi^-\pi^+$ and $\bar{B}^0\rightarrow \bar{K}^*(892)f_0(500)\rightarrow K^-\pi^+\pi^-\pi^+$ channels, we predict $\mathcal{A_{CP}}(\bar{B}^0\rightarrow K^-\pi^+\pi^-\pi^+)\in[0.15,0.28]$ and $\mathcal{B}(\bar{B}^0\rightarrow K^-\pi^+\pi^-\pi^+)\in[1.73,5.10]\times10^{-7}$, respectively, which shows that this two channels' interference mechanism can induce the localized $CP$ violation to this four-body decay. Meanwhile, within the two quark model framework for the scalar mesons $f_0(500)$ and $\bar{K}_0^*(700)$, we calculate the direct CP violations and branching fractions of the $\bar{B}^0\rightarrow \bar{K}_0^*(700)\rho^0(770)$ and $\bar{B}^0\rightarrow \bar{K}^*(892)f_0(500)$ decays, respectively. The corresponding results are $\mathcal{A_{CP}}(\bar{B}^0\rightarrow \bar{K}_0^*(700)\rho^0(770)) \in [0.20, 0.36]$, $\mathcal{A_{CP}}(\bar{B}^0\rightarrow \bar{K}^*(892)f_0(500))\in [0.08, 0.12]$, $\mathcal{B} (\bar{B}^0\rightarrow \bar{K}_0^*(700)\rho^0(770)\in [6.76, 18.93]\times10^{-8}$ and $\mathcal{B} (\bar{B}^0\rightarrow \bar{K}^*(892)f_0(500))\in [2.66, 4.80]\times10^{-6}$, respectively, indicating the $CP$ violations of these two two-body decays are both positive and the branching fractions are quite different. These studies provide a new way to investigate the aforementioned four-body decay and could be helpful in clarifying the configuration of the structure of light scalar meson.
\end{abstract}


\maketitle

\section{Introduction}
Charge-Parity ($CP$) violation is one of the most fundamental and important properties of the weak interaction. Nonleptonic decays of hadrons containing a heavy quark play an important role in testing the Standard Model (SM) picture for the $CP$ violation mechanism in flavor physics, improving our understanding of nonperturbative and perturbative QCD and exploring new physics beyond the SM. $CP$ violation is related to the weak complex phase in the Cabibbo-Kobayashi-Maskawa (CKM) matrix, which describes the mixing of different generations of quarks \cite{Cabibbo:1963yz, Kobayashi:1973fv}. Besides the weak phase, a large strong phase is usually also needed for a large $CP$ violation. Generally, this strong phase is provided by QCD loop corrections and some phenomenological models.

Recently, more attentions have been focused on the studies of the two- or three-body heavy meson decays both theoretically and experimentally \cite{Aubert:2008bj,Garmash:2005rv,Aaij:2013sfa,Beneke:2003zv,Cheng:2013dua,Lu:2000em,Xiao:2011tx,Li:2015zra,Chang:2014rla,Zhang:2013oqa}, while for the four-body nonleptonic decays of these heavy mesons there are limited  studies \cite{Cheng:2017qpv,Pavao:2017kcr,Akar:2018zhv}. Because of the complicated phase spaces and relatively smaller branching fractions, four-body decays of heavy mesons are hard to be investigated. However, in the aspect of studying the intermediate resonances, four-body decays of heavy mesons can provide rich information, especially for the unclear compositions of scalar mesons like $f_0(500)$ ($\sigma$), $K^*(700)$ ($\kappa$), $a_0(980)$ and $f_0(980)$. Up to now, the descriptions of the inner structures for the light scalar states are still unclear and even controversial, which could be, for example, $q\bar{q}$, $\bar{q}\bar{q}qq$, meson-meson bound states or even those supplemented with a scalar glueball. Studying four-body decays of heavy mesons in addition to two- or three- body decays can provide useful information for clarifying configurations of light scalar mesons. In fact, with the great development of the large hadron collider beauty (LHCb) and Belle-II experiments, more and more four-body decay modes involving one or two scalar states in the $B$ and $D$ meson decays are expected to be measured with good precision in the future.

As mentioned above, four-body meson decays are generally dominated by intermediate resonances, which means that they proceed through quasi-two-body or quasi-three-body decays.  In our work, we will adopt the quasi-two-body decay mechanism to study the four-body decay $\bar{B}^0\rightarrow K^-\pi^+\pi^-\pi^+$, i.e. $\bar{B}^0\rightarrow \bar{K}_0^*(700)\rho^0(770)\rightarrow K^-\pi^+\pi^-\pi^+$ and $\bar{B}^0\rightarrow \bar{K}^*(892)f_0(500)\rightarrow K^-\pi^+\pi^-\pi^+$, where the light scalars $f_0(500)$ and $K^*(700)$ will be considered as lowest-lying and first excisted $q\bar{q}$ states \cite{Cheng:2005nb}, respectively. We can then explore whether the localized CP violation of the four-body decay $\bar{B}^0\rightarrow K^-\pi^+\pi^-\pi^+$ can be induced by these two channels' interference.

Theoretically, to calculate the hadronic matrix elements of $B$ or $D$ weak decays, some approaches, such as QCD factorization (QCDF) \cite{Beneke:2003zv,Beneke:2001ev}, the perturbative QCD(pQCD) \cite{Keum:2000ph} and the soft-collinear effective theory(SCET) \cite{Bauer:2000ew}, have been fully developed and extensively employed in recent years. Unfortunately, in the collinear factorization approximation, the calculation of annihilation corrections always suffers from the end-point divergence. In the QCDF approach, such divergence is usually parameterized in a model-independent manner \cite{Beneke:2003zv,Beneke:2001ev} and will be explicitly expressed in Sect. ${\mathrm{\uppercase\expandafter{\romannumeral2}}}$.

The remainder of this paper is organized as follows. In Sect. ${\mathrm{\uppercase\expandafter{\romannumeral2}}}$, we present our theoretical framework. The numerical results are given in Sect. ${\mathrm{\uppercase\expandafter{\romannumeral3}}}$ and we summarize our work in Sect ${\mathrm{\uppercase\expandafter{\romannumeral4}}}$. Appendix A recapitulates explicit expressions of hard spectator-scattering and weak annihilation amplitudes. The factorizable amplitudes of two-body decays are summarized in Appendix B. Related theoretical parameters are listed in Appendix C.

\section{THEORETICAL FRAMEWORK}
\subsection{Kinematics of the four-body decay}
\begin{figure}[ht]
\centerline{\includegraphics[width=0.5\textwidth]{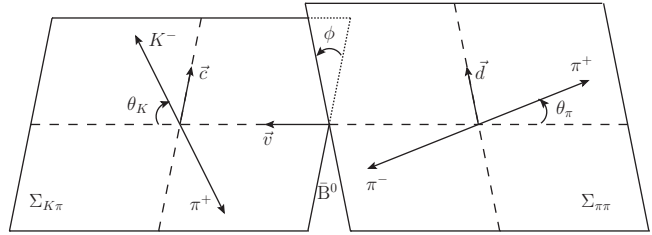}}
\caption{The reference frames and the kinematic variables in the $\bar{B}^0\rightarrow K^-\pi^+\pi^-\pi^+$ decay.}
\label{p3}
\end{figure}

The kinematics of the process $\bar{B}^0\rightarrow K^-(p_1)\pi^+(p_2)\pi^-(p_3)\pi^+(p_4)$ is described in terms of the five variables displayed in Fig. \ref{p3} \cite{Bijnens:1994ie,Kang:2013jaa} in which

\begin{itemize}
\item[(\romannumeral1)] the invariant mass squared of the $K\pi$ system $s_{K\pi}=(p_1+p_2)^2=m_{K\pi}^2$;
\item[(\romannumeral2)] the invariant mass squared of the $\pi\pi$ system $s_{\pi\pi}=(p_3+p_4)^2=m_{\pi\pi}^2$;
\item[(\romannumeral3)] $\theta_\pi$ is the angle of the $\pi^+$ in the $\pi^-\pi^+$ center-of-mass frame $\Sigma_{\pi\pi}$ with respect to the $\pi s$' line of flight in the $\bar{B}^0$ rest frame $\Sigma_{\bar{B}^0}$;
\item[(\romannumeral4)] $\theta_K$ is the angle of the $K^-$ in the $K\pi$ center-of-mass system $\Sigma_{K\pi}$ with respect to the $K\pi$ line of flight in $\Sigma_{\bar{B}^0}$;
\item[(\romannumeral5)] $\phi$ is the angle between the $K\pi$ and $\pi\pi$ planes.
\end{itemize}

The physical ranges are
\begin{equation}\label{Mand}
\begin{split}
4m_{\pi\pi}^2&\leq s_{\pi\pi}\leq(m_{\bar{B}^0}-m_{K\pi})^2,\\
(m_K+m_\pi)^2&\leq s_{K\pi}\leq(m_{\bar{B}^0}-\sqrt{s_{\pi\pi}})^2,\\
0&\leq\theta_\pi,\theta_K\leq\pi,\quad 0\leq\phi\leq2\pi.\\
 \end{split}
 \end{equation}

 We consider the localize $CP$ violation of the $\bar{B}^0\rightarrow K^-(p_1)\pi^+(p_2)\pi^-(p_3)\pi^+(p_4)$ decay when the invariant mass of $\pi\pi$ is near the masses of $\rho^0(770)$ and $f_0(500)$, and the invariant mass of $K\pi$ is near the masses of $\bar{K}_0^*(700)$ and $\bar{K}^*(892)$, respectively. We adopt
\begin{equation}\label{Mand1}
\begin{split}
\bigg(m_{\rho^0(770)}-\frac{\Gamma_{\rho^0(770)}}{2}\bigg)^2&\leq s_{\pi\pi}\leq\bigg(m_{\rho^0(770)}+\frac{\Gamma_{\rho^0(770)}}{2}\bigg)^2,\\
\bigg(m_{\bar{K}_0^*(700)}-\frac{\Gamma_{\bar{K}_0^*(700)}}{2}\bigg)^2&\leq s_{K\pi}\leq\bigg( m_{\bar{K}_0^*(700)}+\frac{\Gamma_{\bar{K}_0^*(700)}}{2}\bigg)^2,\\
 \end{split}
 \end{equation}
for $\bar{B}^0\rightarrow \bar{K}_0^*(700)\rho^0(770)\rightarrow K^-\pi^+\pi^-\pi^+$ decay, and

 \begin{equation}\label{Mand2}
\begin{split}
\bigg(m_{f_0(500)}-\frac{\Gamma_{f_0(500)}}{2}\bigg)^2&\leq s_{\pi\pi}\leq\bigg(m_{f_0(500)}+\frac{\Gamma_{f_0(500)}}{2}\bigg)^2,\\
\bigg(m_{\bar{K}^*(892)}-\frac{\Gamma_{\bar{K}^*(892)}}{2}\bigg)^2&\leq s_{K\pi}\leq \bigg(m_{\bar{K}^*(892)}+\frac{\Gamma_{\bar{K}^*(892)}}{2}\bigg)^2,\\
 \end{split}
 \end{equation}
for $\bar{B}^0\rightarrow \bar{K}^*(892)f_0(500)\rightarrow K^-\pi^+\pi^-\pi^+$ decay, respectively. In Eqs. (\ref{Mand1}) and (\ref{Mand2}), $m_{\rho^0(770)}$, $m_{f_0(500)}$, $m_{\bar{K}_0^*(700)}$ and $m_{\bar{K}^*(892)}$ are the masses of $\rho^0(770)$, $f_0(500)$, $\bar{K}_0^*(700)$ and $\bar{K}^*(892)$ mesons, respectively. $\Gamma_{\rho^0(770)}$, $\Gamma_{f_0(500)}$, $\Gamma_{\bar{K}_0^*(700)}$ and $\Gamma_{\bar{K}^*(892)}$ are the widthes of the corresponding mesons, respectively.

Instead of the individual momenta $p_1$, $p_2$, $p_3$, $p_4$, it is more convenient to use the following kinematic variables
\begin{equation}\label{kinvar}
\begin{split}
 P&=p_1+p_2,\quad Q=p_1-p_2,\\
 L&=p_3+p_4,\quad N=p_3-p_4.\\
 \end{split}
 \end{equation}
It follows that
\begin{equation}\label{kinvar}
\begin{split}
 P^2&=s_{K\pi},\quad Q^2=2(p_K^2+p_\pi^2)-s_{K\pi},\quad L^2=s_{\pi\pi},\\
 P\cdot L&=\frac{1}{2}(m_{\bar{B}^0}^2-s_{K\pi}-s_{\pi\pi}),\quad P\cdot N=X\cos\theta_1.\\
 \end{split}
 \end{equation}
where the function $X$ is defined as
\begin{equation}\label{X}
\begin{split}
 X(s_{K\pi},s_{\pi\pi})&=\bigg[(P\cdot L)^2-s_{K\pi}s_{\pi\pi}\bigg]^{1/2}=\frac{1}{2}\lambda^{1/2}(m_{\bar{B}^0}^2,s_{K\pi},s_{\pi\pi}),\\
\lambda(x,y,z)&=(x-y-z)^2-4yz.\\
 \end{split}
 \end{equation}
\subsection{B decay in QCD factorization}
The effective weak Hamiltonian for nonleptonic $B$ weak decays is \cite{Beneke:2003zv}
 \begin{equation}\label{Hamiltonian}
 \mathcal{H}_{eff}=\frac{G_F}{\sqrt{2}}\bigg[\sum_{p=u,c}\sum_{D=d,s}\lambda_{p}^{(D)}(c_1O_1^p+c_2O_2^p+\sum_{i=3}^{10}c_iO_i+c_{7\gamma}O_{7\gamma}+c_{8g}O_{8g})\bigg]+h.c.,
 \end{equation}
 where $G_F$ represents the Fermi constant, $\lambda_p^{(D)}=V_{pb}V_{pD}^*$, $V_{pb}$ and $V_{pD}$ are the CKM matrix elements, $c_i (i=1-10,7\gamma,8g)$ are Wilson coefficients, $O_{1,2}^p$ are the tree level operators, $O_{3-6}$ are the QCD penguin operators, $O_{7-8}$ arise from electroweak penguin diagrams, and $O_{7\gamma}$ and$O_{8g}$ are the electromagnetic and chromomagnetic dipole operators, respectively.

With the effective Hamiltonian in Eq. (\ref{Hamiltonian}), the QCDF method has been fully developed and extensively employed to calculate the hadronic two-body B decays. The spectator scattering and annihilation amplitudes are expressed with the convolution of scattering functions and the light-cone wave functions of the participating mesons \cite{Beneke:2003zv}. The explicit expressions for the basic building blocks of the spectator scattering and annihilation amplitudes have been given in Ref. \cite{Beneke:2003zv}, which are also listed in Appendix A for convenience. The annihilation contributions $A_n^{i,f}$ ($n=1,2,3$) can be simplified to \cite{Cheng:2007st}:
\begin{equation}\label{APS}
\begin{split}
A_1^i(VS)&\approx6\pi\alpha_s\bigg\{3\mu_S\bigg[B_1(3X_A+4-\pi^2)+B_3(10X_A+\frac{23}{18}-\frac{10}{3}\pi^2)\bigg]-r_\chi^Sr_\chi^VX_A(X_A-2)\bigg\},\\
A_2^i(VS)&\approx 6\pi\alpha_s\bigg\{3\mu_S\bigg[B_1(X_A+29-3\pi^2)+B_3(X_A+\frac{2956}{9}-\frac{100}{3}\pi^2)\bigg]-r_\chi^Sr_\chi^VX_A(X_A-2)\bigg\},\\
A_3^i(VS) &\approx 6\pi\alpha_s\bigg\{-r^V_\chi\mu_S\bigg[9B_1(X_A^2-4X_A-4+\pi^2)+10B_3(3X_A^2-19X_A+\frac{61}{6}+3\pi^2)\bigg]\\
&-r_\chi^S(X_A^2-2X_A+\frac{\pi^2}{3})\bigg\},\\
A_3^f(VS) &\approx 6\pi\alpha_s\bigg\{-3r^V_\chi\mu_S(X_A-2)\bigg[B_1(6X_A-11)+B_3(20X_A-\frac{187}{3})\bigg]+r_\chi^SX_A(2X_A-1)\bigg\},\\
A_1^f(VS)&=A_2^f(VS)=0,
\end{split}
\end{equation}
for $M_1M_2=VS$, and
\begin{equation}
A_1^i(SV)=-A_2^i(SV),\quad A_2^i(SV)=-A_1^i(SV),\quad A_3^i(SV)=A_3^i(VS),\quad A_3^f(SV)=-A_3^f(VS),
\end{equation}
for $M_1M_2=SV$, where the superscripts $i$ and $f$ refer to gluon emission from the initial and final state quarks, respectively. The model-dependent parameter $X_A$ is used to estimate the end point contributions, and expressed as
\begin{equation}
 X_A=(1+\rho_A e^{i\phi_A})\ln\frac{m_B}{\Lambda_h},
\end{equation}
with $\Lambda_h$ being a typical scale of order 500 $\mathrm{MeV}$, $\rho_A$ an unknown real parameter and $\phi_A$ the free strong phase in the range $[0,2\pi]$. For the spectator scattering contributions, the calculation of twist-3 distribution amplitudes also suffers from the end point divergence, which is usually dealt with in the same manner as in Eq. (9) and labeled by $X_H$. In our work, when dealing with the end-point divergences from the hard spectator scattering and weak annihilation contributions, we will follow the assumption $X_H=X_A$ for the $B$ two-body decays \cite{Cheng:2005nb}.  Moreover, a quantity $\lambda_B$ is introduced to parametrize the integral over the $B$ meson distribution amplitude through \cite{Beneke:2003zv}
\begin{equation}
 \int_0^1\frac{d\rho}{\rho}\Phi_B(\rho)\equiv\frac{m_B}{\lambda_B}.\\
\end{equation}

\subsection{Four-body decay amplitudes and localized CP violation}
For the $\bar{B}^0\rightarrow K^-\pi^+\pi^-\pi^+$ decay, we consider the contributions from $\bar{B}^0\rightarrow \bar{K}_0^*(700)\rho^0(770)\rightarrow K^-\pi^+\pi^-\pi^+$ and $\bar{B}^0\rightarrow \bar{K}^*(892)f_0(500)\rightarrow K^-\pi^+\pi^-\pi^+$ channels. For convenience, $f_0(500)$, $\rho^0(770)$, $\bar{K}_0^*(700)$ and $\bar{K}^*(892)$ mesons will be denoted as $\sigma$, $\rho$, $\bar{\kappa}$ and $\bar{K}^*$, respectively. The amplitudes of these two channels are
\begin{equation} \label{HKrho1}
\begin{split}
\mathcal{M}(\bar{B}^0\rightarrow \bar{\kappa}\rho\rightarrow K^-\pi^+\pi^+\pi^- )=\frac{\langle \bar{\kappa}\rho|\mathcal{H}_{eff}|\bar{B}^0\rangle \langle K^-\pi^+|\mathcal{H}_{\bar{\kappa}\pi^+\pi^-}|\bar{\kappa}\rangle \langle \pi^-\pi^+|\mathcal{H}_{\rho \pi^-\pi^+}|\rho\rangle}{S_{\bar{\kappa}}S_{\rho}},
 \end{split}
 \end{equation}
and
\begin{equation} \label{HKrho2}
\begin{split}
\mathcal{M}(\bar{B}^0\rightarrow \bar{K}^*\sigma\rightarrow K^-\pi^+\pi^+\pi^-)=\frac{\langle \bar{K}^*\sigma|\mathcal{H}_{eff}|\bar{B}^0\rangle \langle K^-\pi^+|\mathcal{H}_{\bar{K}^*\pi^+\pi^-}|\bar{K}^*\rangle \langle \pi^-\pi^+|\mathcal{H}_{\sigma \pi^-\pi^+}|\sigma\rangle}{S_{\bar{K}^*}S_{\sigma}},
 \end{split}
 \end{equation}
respectively, where $\mathcal{H}_{\rho\pi^+\pi^-}$, $\mathcal{H}_{\sigma\pi^+\pi^-}$, $\mathcal{H}_{\bar{\kappa}K^-\pi^+}$ and $\mathcal{H}_{\bar{K}^*K^-\pi^+}$ are strong Hamiltonians for $\rho\rightarrow\pi^-\pi^+$, $\sigma\rightarrow\pi^-\pi^+$, $\bar{\kappa}\rightarrow K^-\pi^+$ and $\bar{K}^*\rightarrow K^-\pi^+$ decays, respectively. $S_{\bar{\kappa}}$, $S_{\rho}$, $S_{\bar{K}^*}$ and $S_{\sigma}$ are the reciprocal of the propagators of the corresponding mesons. We shall adopt the
Breit-Wigner function and the Bugg model \cite{Bugg:2006gc,Qi:2018syl} to deal with the distributions of the first three mesons ($\bar{\kappa}$, $\rho$ and $\bar{K}^*$) and $\sigma$ meson, respectively.

Within the QCDF framework in Ref. \cite{Beneke:2003zv}, we can get the decay amplitudes of $\bar{B}^0\rightarrow \bar{\kappa}\rho, \bar{K}^*\sigma $ which have been listed in Appendix B. Combining Eqs. (\ref{amplitude12}) and (\ref{HKrho1}), (\ref{amplitude22}) and (\ref{HKrho2}), respectively, the amplitudes of $\bar{B}^0\rightarrow \bar{\kappa}\rho\rightarrow K^-\pi^+\pi^-\pi^+ $ and $\bar{B}^0\rightarrow \bar{K}^*\sigma\rightarrow K^-\pi^+\pi^-\pi^+$ channels can be written as
\begin{eqnarray}\label{amplitude11}
\begin{split}
\mathcal{M}(\bar{B}^0\rightarrow \bar{\kappa}\rho\rightarrow K^-\pi^+\pi^+\pi^- )&=\frac{iG_Fg_{\bar{\kappa}K\pi}g_{\rho\pi\pi}\varepsilon_{\bar{\kappa}}\cdot(p_{K^-}-p_{\pi^+})\varepsilon_{\rho}\cdot(p_{\pi^-}-p_{\pi^+})}{S_{\bar{\kappa}}S_\rho }\sum_{p=u,c}\lambda_p^{(s)}\\
&\times\bigg\{\bigg[\delta_{pu}\alpha_2(\bar{\kappa}\rho)
+\frac{3}{2}\alpha_{3,EW}^p(\bar{\kappa}\rho)\bigg]
\times 2f_\rho m_\rho \varepsilon_\rho^*\cdot p_BF_1^{\bar{B}^0 \bar{\kappa}}(m_{\rho}^2)\\
&+\bigg[\alpha_4^p(\rho\bar{\kappa})-\frac{1}{2}\alpha_{4,EW}^p(\rho\bar{\kappa})\bigg]
\times2\bar{f}_{\bar{\kappa}}m_\rho \varepsilon_\rho^*\cdot p_BA_0^{\bar{B}^0\rho}(m_{\bar{\kappa}}^2)\\
&+\bigg[b_3^p(\rho\bar{\kappa})-\frac{1}{2}b_{3,EW}^p(\rho\bar{\kappa})\bigg]
\times2m_{\rho}f_{\bar{B}^0}f_\rho \bar{f}_{\bar{\kappa}}\bigg\},\\
\end{split}
\end{eqnarray}
and

\begin{eqnarray}\label{amplitude21}
\begin{split}
\mathcal{M}(\bar{B}^0\rightarrow \bar{K}^*\sigma\rightarrow K^-\pi^+\pi^+\pi^-)&=-\frac{iG_Fg_{\bar{K}^*K\pi}g_{\sigma\pi\pi}}{S_{\bar{K}^*}S_\sigma }\sum_{p=u,c}\lambda_p^{(s)}\bigg\{\bigg[\frac{1}{\sqrt{2}}\delta_{pu}\alpha_2(\bar{K}^*\sigma)+\sqrt{2}\alpha_3^p(\bar{K}^*\sigma)\\
&+\frac{1}{2\sqrt{2}}\alpha_{3,EW}^p(\bar{K}^*\sigma)\bigg]\times 2\bar{f}_{\sigma^s}A_0^{\bar{B}^0 \bar{K}^*}(m_\sigma^2)+\bigg[\alpha_3^p(\bar{K}^*\sigma)+\alpha_4^p(\bar{K}^*\sigma)\\
&-\frac{1}{2}\alpha_{3,EW}^p(\bar{K}^*\sigma)-\frac{1}{2}\alpha_{4,EW}^p(\bar{K}^*\sigma)\bigg]\times2\bar{f}_{\sigma^s}A_0^{\bar{B}^0\bar{K}^*}(m_\sigma^2)+\bigg[\frac{1}{2\sqrt{2}}\alpha_{4,EW}^p(\sigma\bar{K}^*)\\
&-\frac{1}{\sqrt{2}}\alpha_4^p(\sigma\bar{K}^*)\bigg]\times2f_{\bar{K}^*}
F_1^{\bar{B}^0\sigma}(m_{\bar{K}^*}^2)-\bigg[b_3^p(\bar{K}^*\sigma)+b_{3,EW}^p(\bar{K}^*\sigma)\bigg]\\
&\times m_{\bar{K}^*}f_{\bar{B}^0}f_{\bar{K}^*}\bar{f}_\sigma^s/(m_{\bar{B}^0}p_c)+\bigg[\frac{1}{\sqrt{2}}b_3^p(\sigma\bar{K}^*)-\frac{1}{2\sqrt{2}}b_{3,EW}^p(\sigma\bar{K}^*)\bigg]\\
&\times m_{\bar{K}^*}f_{\bar{B}^0}f_{\bar{K}^*}\bar{f}_\sigma^n/(m_{\bar{B}^0}p_c)\bigg\},\\
\end{split}
\end{eqnarray}
respectively, where $g_{\bar{\kappa}K\pi}$, $g_{\rho\pi\pi}$, $g_{\bar{K}^*K\pi}$, $g_{\sigma \pi\pi}$ are the strong coupling constants of the corresponding decays, which are listed in Eq. (\ref{gSV}), $F_1^{\bar{B}^0 \bar{\kappa}}(m_{\rho}^2)$, $A_0^{\bar{B}^0\rho}(m_{\bar{\kappa}}^2)$, $A_0^{\bar{B}^0 \bar{K}^*}(m_\sigma^2)$ and $F_1^{\bar{B}^0\sigma}(m_{\bar{K}^*}^2)$ are form factors for $\bar{B}^0$ to $\bar{\kappa}$, $\rho$, $\bar{K}^*$ and $\sigma$ transitions, respectively, $f_\rho$,  $\bar{f}_{\bar{\kappa}}$, $f_{\bar{B}^0}$ and $f_{\bar{K}^*}$ are decay constants of $\rho$, $\bar{\kappa}$, $\bar{B}^0$ and $bar{K}^*$ mesons, respectively, $\bar{f}_{\sigma^s}$ and $\bar{f}_\sigma^n$ are decay constants of $\sigma$ coming from the up and strange quark components, respectively.

The total decay amplitude of the $\bar{B}^0\rightarrow K^-\pi^+\pi^+\pi^-$ including both $\bar{B}^0\rightarrow \bar{\kappa}\rho\rightarrow K^-\pi^+\pi^+\pi^-$ and $\bar{B}^0\rightarrow \bar{K}^*\sigma\rightarrow K^-\pi^+\pi^+\pi^-$ channels can be written as
\begin{equation}\label{A}
\mathcal{M}=\mathcal{M}(\bar{B}^0\rightarrow \bar{\kappa}\rho\rightarrow K^-\pi^+\pi^+\pi^- )+\mathcal{M}(\bar{B}^0\rightarrow \bar{K}^*\sigma\rightarrow K^-\pi^+\pi^+\pi^-).
 \end{equation}

The differential CP asymmetry parameter can be defined as
 \begin{equation}\label{CP asymmetry parameter}
\mathcal{A_{CP}}=\frac{|\mathcal{M}|^2-|\bar{\mathcal{M}}|^2}{|\mathcal{M}|^2+|\bar{\mathcal{M}}|^2}.
 \end{equation}

The localized integration $CP$ asymmetry can be measured by experiments and takes the following form:
  \begin{equation}\label{localized CP}
\mathcal{A_{CP}}=\frac{\int d\Omega(|\mathcal{M}|^2-|\bar{\mathcal{M}}|^2)}{\int d\Omega(|\mathcal{M}|^2+|\bar{\mathcal{M}}|^2)},
 \end{equation}
where $\Omega$ represents the phase space given in Eqs. (\ref{Mand1}) and (\ref{Mand2}) with $d\Omega=ds_{\pi\pi}ds_{K\pi}dcos\theta_\pi dcos\theta_Kd\phi$.

As for the decay rate, one has \cite{Cheng:2017qpv}
\begin{equation}
d^5\Gamma=\frac{1}{4(4\pi)^6m_{\bar{B}^0}^3}\sigma(s_{\pi\pi})X(s_{\pi\pi},s_{K\pi})\sum_{\mathrm{spins}}|\mathcal{M}|^2d\Omega,
\end{equation}
with
\begin{equation}
\sigma(s_{\pi\pi})=\sqrt{1-4m_\pi^2/s_{\pi\pi}}.
\end{equation}
This leads to the branching fraction
\begin{equation}\label{B}
\mathcal{B}=\frac{1}{\Gamma_{\bar{B}^0}}\int d^5\Gamma,
\end{equation}
where $\Gamma_{\bar{B}^0}$ is the decay width of the $\bar{B}^0$ meson.
\section{NUMERICAL RESULTS}
Within the QCDF approach, we get the amplitudes of the two-body decays $\bar{B}^0\rightarrow \bar{\kappa}\rho$ and $\bar{B}^0\rightarrow \bar{K}^*\sigma$, where the light scalar $\sigma$ and $\bar{\kappa}$ mesons are considered as the lowest-lying and first excisted $q\bar{q}$ states \cite{Cheng:2005nb}, respectively. As for the parameters for the end-point divergences, we take $\rho_{H(A)}\leq0.5$ and arbitrary strong phases $\phi_{A(H)}$. All the form factors are evaluated at $q^2=0$ due to the smallness of $m_{\rho}^2$, $m_{\bar{\kappa}}^2$, $m_\sigma^2$ and $m_{\bar{K}^*}^2$ compared with $m_{\bar{B}^0}^2$. We also
simply set $F^{\bar{B}^0\rightarrow\kappa}(0)=0.3$ and assign its uncertainty to be $\pm0.1$. With the given parameters, we obtain the $CP$ violations and branching fractions of the $\bar{B}^0\rightarrow \bar{\kappa}\rho$ and $\bar{B}^0\rightarrow \bar{K}^*\sigma$ decays substituting Eqs. (\ref{amplitude12}), (\ref{amplitude22}) into (\ref{CP asymmetry parameter}), respectively. The results are $\mathcal{A_{CP}}(\bar{B}^0\rightarrow \bar{\kappa}\rho) \in [0.20, 0.36]$, $\mathcal{A_{CP}}(\bar{B}^0\rightarrow \bar{K}^*\sigma)\in [0.08, 0.12]$, $\mathcal{B} (\bar{B}^0\rightarrow \bar{\kappa}\rho)\in [6.76, 18.93]\times10^{-8}$ and $\mathcal{B} (\bar{B}^0\rightarrow \bar{K}^*\sigma)\in [2.66, 4.80]\times10^{-6}$, respectively. Obviously, the $CP$ violations of these two-body decays are both positive, with the $CP$ violation in $\bar{B}^0\rightarrow \bar{K}^*\sigma$ decay being smaller than that in $\bar{B}^0\rightarrow \bar{\kappa}\rho$. The magnitudes of the branching fractions in these two-body decays are different with the former being about two orders smaller than the latter. When dealing with the the four-body decay $\bar{B}^0\rightarrow K^-\pi^+\pi^-\pi^+$, we adopt $\mathcal{B}(\rho\rightarrow \pi^-\pi^+)\approx 1$, $\mathcal{B}(\sigma\rightarrow \pi^-\pi^+)\approx \frac{2}{3}$, $\mathcal{B}(\bar{K}^*\rightarrow K^-\pi^+)\approx 1$, $\mathcal{B}(\bar{\kappa}\rightarrow K^-\pi^+)\approx \frac{2}{3}$. Then, substituting Eq. (\ref{A}) into (\ref{localized CP}) and (\ref{B}), respectively, we get the localized $CP$ violation and branching fraction of the four-body decay $\bar{B}^0\rightarrow K^-\pi^+\pi^-\pi^+$, with the results
 $\mathcal{A_{CP}}(\bar{B}^0\rightarrow K^-\pi^+\pi^-\pi^+)\in[0.15,0.28]$ and $\mathcal{B}(\bar{B}^0\rightarrow K^-\pi^+\pi^-\pi^+)\in[1.73,5.10]\times10^{-7}$. It is obvious that the sign of the localized $CP$ violation of $\bar{B}^0\rightarrow K^-\pi^+\pi^-\pi^+$ is positive when the invariant masses of $\pi\pi$ and $K\pi$ are near the masses of $\rho$ ($\sigma$) and $\bar{\kappa}$ ($\bar{K}^*$), respectively. This indicates that the interference of
 $\bar{B}^0\rightarrow \bar{\kappa}\rho\rightarrow K^-\pi^+\pi^-\pi^+$ and $\bar{B}^0\rightarrow \bar{K}^*\sigma\rightarrow K^-\pi^+\pi^-\pi^+$ channels can induce the localized $CP$ violation to the four-body decay $\bar{B}^0\rightarrow K^-\pi^+\pi^-\pi^+$. Our theoretical results shown here are predictions for ongoing experiments at LHCb and Belle-II.

\section{SUMMARY}
By studying the quasi-two-body decays within the QCDF approach, we predicted the localized $CP$ violation and branching fraction of the four-body decay $\bar{B}^0\rightarrow K^-\pi^+\pi^-\pi^+$ due to the interference of the two channels $\bar{B}^0\rightarrow \bar{K}_0^*(700)\rho^0(770)(\rightarrow\bar{\kappa}\rho)\rightarrow K^-\pi^+\pi^-\pi^+$ and $\bar{B}^0\rightarrow \bar{K}^*(892)f_0(500)(\rightarrow\bar{K}^*\sigma)\rightarrow K^-\pi^+\pi^-\pi^+$, with the results $\mathcal{A_{CP}}(\bar{B}^0\rightarrow K^-\pi^+\pi^-\pi^+)\in[0.15,0.28]$ and $\mathcal{B}(\bar{B}^0\rightarrow K^-\pi^+\pi^-\pi^+)\in[1.73,5.10]\times10^{-7}$. It is obvious that the sign of the localized $CP$ violation of $\bar{B}^0\rightarrow K^-\pi^+\pi^-\pi^+$ is positive. In the two quark model for the scalar mesons, we also obtained the $CP$ violations and branching fractions of the two-body decays $\bar{B}^0\rightarrow \bar{K}_0^*(700)\rho^0(770)$ and $\bar{B}^0\rightarrow \bar{K}^*(892)f_0(500)$ as $\mathcal{A_{CP}}(\bar{B}^0\rightarrow \bar{K}_0^*(700)\rho^0(770)) \in [0.20, 0.36]$, $\mathcal{A_{CP}}(\bar{B}^0\rightarrow \bar{K}^*(892)f_0(500))\in [0.08, 0.12]$, $\mathcal{B} (\bar{B}^0\rightarrow \bar{K}_0^*(700)\rho^0(770)\in [6.76, 18.93]\times10^{-8}$ and $\mathcal{B} (\bar{B}^0\rightarrow \bar{K}^*(892)f_0(500))\in [2.66, 4.80]\times10^{-6}$, respectively. Obviously, the $CP$ violations of these two-body decays are both positive, and the $CP$ violation in $\bar{B}^0\rightarrow \bar{K}^*(892)f_0(500)$ is smaller than that in $\bar{B}^0\rightarrow \bar{K}_0^*(700)\rho^0(770)$. Futhermore, the branching fractions in these two body decays are quite different, with the former being two orders smaller than the latter. Our results will be tested by the precise data from future LHCb and Belle-II experiments. In the present work, we assumed that $f_0(500)$ and $\bar{K}_0^*(700)$ are dominated by the $q\bar{q}$ configuration. Possible other structures of $f_0(500)$ and $\bar{K}_0^*(700)$ could affect the results in our interference model which will need further investigation.

\begin{appendix}
\section{Explicit expressions of hard spectator-scattering and weak annihilation amplitudes}

 For the hard spectator terms, we obtain \cite{Cheng:2007st}
 \begin{equation}
  \begin{split}
 H_i(M_1M_2)=-\frac{f_{\bar{B}^0}f_{M_1}}{D(M_1M_2)}\int_0^1\frac{d\rho}{\rho}\Phi_{\bar{B}^0}(\rho)\int_0^1\frac{d\xi}{\bar{\xi}}\Phi_{M_2}(\xi)
 \int_0^1\frac{d\eta}{\bar{\eta}}[\pm\Phi_{M_1}(\eta)+r_\chi^{M_1}\frac{\bar{\xi}}{\xi}\Phi_{m_1}(\eta)],
  \end{split}
  \end{equation}
 for $i=1-4,9,10$, where the upper sign is for $M_1=V$ and the lower sign for $M_1=S$,
 \begin{equation}
  \begin{split}
 H_i(M_1M_2)=-\frac{f_{\bar{B}^0}f_{M_1}}{D(M_1M_2)}\int_0^1\frac{d\rho}{\rho}\Phi_{\bar{B}^0}(\rho)\int_0^1\frac{d\xi}{\xi}\Phi_{M_2}(\xi)
 \int_0^1\frac{d\eta}{\bar{\eta}}[\pm\Phi_{M_1}(\eta)+r_\chi^{M_1}\frac{\xi}{\bar{\xi}}\Phi_{m_1}(\eta)],
  \end{split}
  \end{equation}
 for $i=5,7$ and $H_i=0$ for $i=6,8$, $\bar{\xi}=1-\xi$ and $\bar{\eta}=1-\eta$, $\Phi_M(\Phi_m)$ is the twist-2 (twist-3) light-cone distribution amplitude of the meson $M$, and
 \begin{equation}
 D(SV)=F_1^{\bar{B}^0S}(0)m_{\bar{B}^0}^2, \quad D(VS)=A_0^{\bar{B}^0V}(0)m_{\bar{B}^0}^2,
 \end{equation}
 and $r_\chi^{M_i}$ (i=1,2) are ``chirally-enhanced" terms defined as
\begin{equation}\label{D}
\begin{split}
 r_\chi^V(\mu)=\frac{2m_V}{m_b(\mu)}\frac{f_V^\perp(\mu)}{f_V},\\
 \bar{r}_\chi^S(\mu)=\frac{2m_S}{m_b(\mu)}.\\
\end{split}
\end{equation}

With the asymptotic light-cone distribution amplitudes, the building blocks for the annihilation amplitudes are given by \cite{Cheng:2007st}
\begin{equation}\label{Ai}
\begin{split}
 A_1^i&=\pi\alpha_s\int_0^1 dx dy\begin{cases}
 \bigg(\Phi_{V}(x)\Phi_{S}(y)\bigg[\frac{1}{x(1-\bar{x}y)}+\frac{1}{x\bar{y}^2}\bigg]+r_\chi^{V}r_\chi^{S} \Phi_{v}(x)\Phi_{S}^s(y)\frac{2}{x\bar{y}}\bigg),\quad \text{for $M_1M_2=VS,$}\\
 \bigg(\Phi_{S}(x)\Phi_{V}(y)\bigg[\frac{1}{x(1-\bar{x}y)}+\frac{1}{x\bar{y}^2}\bigg]+r_\chi^{V}r_\chi^{S} \Phi_{S}^s(x)\Phi_{v}(y)\frac{2}{x\bar{y}}\bigg),\quad \text{for $M_1M_2=SV,$}\\
 \end{cases}\\
 A_2^i&=\pi\alpha_s\int_0^1 dx dy\begin{cases}
 \bigg(\Phi_{V}(x)\Phi_{S}(y)\bigg[\frac{1}{\bar{y}(1-\bar{x}y)}+\frac{1}{x^2\bar{y}}\bigg]+r_\chi^{V}r_\chi^{S} \Phi_{v}(x)\Phi_{S}^s(y)\frac{2}{x\bar{y}}\bigg),\quad \text{for $M_1M_2=VS,$}\\
 \bigg(\Phi_{S}(x)\Phi_{V}(y)\bigg[\frac{1}{\bar{y}(1-\bar{x}y)}+\frac{1}{x^2\bar{y}}\bigg]+r_\chi^{V}r_\chi^{S} \Phi_{S}^s(x)\Phi_{v}(y)\frac{2}{x\bar{y}}\bigg),\quad \text{for $M_1M_2=SV,$}\\
 \end{cases}\\
 A_3^i&=\pi\alpha_s\int_0^1 dx dy\begin{cases}
 \bigg(r_\chi^{V}\Phi_{v}(x)\Phi_{S}(y)\frac{2\bar{x}}{x\bar{y}(1-\bar{x}y)}-r_\chi^{S} \Phi_{V}(x)\Phi_S^s(y)\frac{2y}{x\bar{y}(1-\bar{x}y)}\bigg),\quad \text{for $M_1M_2=VS,$}\\
\bigg(-r_\chi^{S}\Phi_{S}^s(x)\Phi_{V}(y)\frac{2\bar{x}}{x\bar{y}(1-\bar{x}y)}+r_\chi^{V} \Phi_{S}(x)\Phi_v(y)\frac{2y}{x\bar{y}(1-\bar{x}y)}\bigg),\quad \text{for $M_1M_2=SV,$}\\
 \end{cases}\\
 A_3^f&=\pi\alpha_s\int_0^1 dx dy\begin{cases}
 \bigg(r_\chi^{V}\Phi_{v}(x)\Phi_{S}(y)\frac{2(1+\bar{y})}{x\bar{y}^2}+r_\chi^{S} \Phi_{V}(x)\Phi_S^s(y)\frac{2(1+x)}{x^2\bar{y}}\bigg),\quad \text{for $M_1M_2=VS,$}\\
 \bigg(-r_\chi^{V}\Phi_{S}^s(x)\Phi_{V}(y)\frac{2(1+\bar{y})}{x\bar{y}^2}-r_\chi^{V} \Phi_{S}(x)\Phi_v(y)\frac{2(1+x)}{x^2\bar{y}}\bigg),\quad \text{for $M_1M_2=SV,$}\\
 \end{cases}\\
 A_1^f&=A_2^f=0.\\
 \end{split}
  \end{equation}

\section{THE AMPLITUDES OF $\bar{B}^0\rightarrow \bar{K}_0^*\rho^0$ AND $\bar{B}^0\rightarrow \bar{K}^*\sigma$ DECAYS}
With the conventions in Ref. [11], we obtain the decay amplitudes for $\bar{B}^0\rightarrow \bar{K}_0^*\rho^0, \bar{K}^*\sigma $ decays within the QCDF framework which have the following forms:
\begin{equation}\label{amplitude12}
\begin{split}
\mathcal{M}(\bar{B}^0\rightarrow \bar{\kappa}\rho)&=\frac{iG_F}{2}\sum_{p=u,c}\lambda_p^{(s)}\bigg\{\bigg[\delta_{pu}\alpha_2(\bar{\kappa}\rho)
+\alpha_{3,EW}^p(\bar{\kappa}\rho)\bigg]\times2f_\rho F_1^{\bar{B}^0 \bar{\kappa}}(m_{\rho}^2)m_{\bar{B}^0}p_c+\bigg[\alpha_4^p(\rho\bar{\kappa})-\frac{1}{2}\alpha_{4,EW}^p(\rho\bar{\kappa})\bigg]\\
&\times2\bar{f}_{\bar{\kappa}}A_0^{\bar{B}^0\rho}(m_{\bar{\kappa}}^2)m_{\bar{B}^0}p_c
+\bigg[b_3^p(\rho\bar{\kappa})-\frac{1}{2}b_{3,EW}^p(\rho\bar{\kappa})\bigg]\times f_{\bar{B}^0}f_\rho f_{\bar{\kappa}}\bigg\},\\
\end{split}
\end{equation}

\begin{equation}\label{amplitude22}
\begin{split}
\mathcal{M}(\bar{B}^0\rightarrow \bar{K}^*\sigma)&=-\frac{iG_F}{2}\sum_{p=u,c}\lambda_p^{(s)}\bigg\{\bigg[\frac{1}{\sqrt{2}}\delta_{pu}\alpha_2(\bar{K}^*\sigma)+\sqrt{2}\alpha_3^p(\bar{K}^*\sigma)
+\frac{1}{2\sqrt{2}}\alpha_{3,EW}^p(\bar{K}^*\sigma)\bigg]\\
&\times 2\bar{f}_{\sigma^s}A_0^{\bar{B}^0 \bar{K}^*}(m_\sigma^2)m_{\bar{B}^0}p_c+\bigg[\alpha_3^p(\bar{K}^*\sigma)+\alpha_4^p(\bar{K}^*\sigma)-\frac{1}{2}
\alpha_{3,EW}^p(\bar{K}^*\sigma)-\frac{1}{2}\alpha_{4,EW}^p(\bar{K}^*\sigma)\bigg]\\
&\times2\bar{f}_{\sigma^s}A_0^{\bar{B}^0\bar{K}^*}(m_\sigma^2)m_{\bar{B}^0}p_c+\bigg[\frac{1}{2\sqrt{2}}\alpha_{4,EW}^p(\sigma\bar{K}^*)-\frac{1}{\sqrt{2}}\alpha_4^p(\sigma\bar{K}^*)\bigg]2f_{\bar{K}^*}
F_1^{\bar{B}^0\sigma}(m_{\bar{K}^*}^2)m_{\bar{B}^0}p_c\\
&-\bigg[b_3^p(\bar{K}^*\sigma)+b_{3,EW}^p(\bar{K}^*\sigma)\bigg]f_{\bar{B}^0}f_{\bar{K}^*}\bar{f}_\sigma^s+
\bigg[\frac{1}{\sqrt{2}}b_3^p(\sigma\bar{K}^*)-\frac{1}{2\sqrt{2}}b_{3,EW}^p(\sigma\bar{K}^*)\bigg]f_{\bar{B}^0}f_{\bar{K}^*}\bar{f}_\sigma^n\bigg\}.\\
\end{split}
\end{equation}
\section{THEORETICAL INPUT PARAMETERS}
The predictions obtained in the QCDF  approach depend on many input parameters. The values of the Wolfenstein parameters are taken from Ref. \cite{Agashe:2014kda}: $\bar{\rho}=0.117\pm0.021$, $\bar{\eta}=0.353\pm0.013$.

The effective Wilson coefficients used in our calculations are taken from Ref. \cite{Qi:2018syl}:
\begin{equation}\label{C}
\begin{split}
&C'_1=-0.3125, \quad C'_2=-1.1502, \\
&C'_3=2.120\times10^{-2}+5.174\times10^{-3}i,\quad C'_4=-4.869\times10^{-2}-1.552\times10^{-2}i, \\
&C'_5=1.420\times10^{-2}+5.174\times10^{-3}i,\quad C'_6=-5.792\times10^{-2}-1.552\times10^{-2}i, \\
&C'_7=-8.340\times10^{-5}-9.938\times10^{-5}i,\quad C'_8=3.839\times10^{-4}, \\
&C'_9=-1.017\times10^{-2}-9.938\times10^{-5}i,\quad C'_{10}=1.959\times10^{-3}. \\
\end{split}
\end{equation}

For the masses used in $\bar{B}^0$ decays, we use the following values (in $\mathrm{GeV}$) \cite{Agashe:2014kda}:
\begin{equation}
\begin{split}
m_u&=m_d=0.0035,\quad m_s=0.119, \quad m_b=4.2,\quad m_{\pi^\pm}=0.14,\quad m_{K^+}=0.494,\\
m_{f_0(500)}&=0.50,\quad m_{{\bar{K}_0^*}(700)}=0.824, \quad m_{\rho^0(770)}=0.775,\quad  m_{\bar{K}^*(892)}=0.895,\quad m_{\bar{B}^0}=5.28,\\
\end{split}
\end{equation}
and widthes are (in $\mathrm{GeV}$) \cite{Agashe:2014kda}
\begin{equation}
\begin{split}
\Gamma_{\rho^0(770)}&=0.149,\quad\Gamma_{f_0(500)}=0.5,\quad\Gamma_{\bar{K}_0^*(700)}=0.047,\quad\Gamma_{\bar{K}^*(892)}=0.047.\\
\end{split}
\end{equation}

The strong coupling constants are determined from the measured partial widths through the relations \cite{Cheng:2013dua,Dedonder:2014xpa}
\begin{equation}\label{gSV}
\begin{split}
g_{S M_1M_2}=\sqrt{\frac{8\pi m_S^2}{p_c(S)}\Gamma_{S\rightarrow M_1M_2}},\\
g_{V M_1M_2}=\sqrt{\frac{6\pi m_V^2}{p_c(V)^3}\Gamma_{V\rightarrow M_1M_2}},\\
\end{split}
\end{equation}
where $p_c(S,V)$ are the magnitudes of the three momenta of the final state mesons in the rest frame of S and V mesons, respectively.

The following related decay constants (in $\mathrm{GeV}$) are used \cite{Cheng:2010yd,Cheng:2005nb}:
\begin{equation}
\begin{split}
f_{\pi^\pm}&=0.131,\quad f_{\bar{B}^0}=0.21\pm0.02, \quad f_{K^-}=0.156\pm0.007,  \\
 \bar{f}^s_{f_0(500)}&=-0.21\pm0.093,\quad \bar{f}_{f_0(500)}^u=0.4829\pm0.076,\quad \bar{f}_{\bar{K}_0^*(700)}=0.34\pm0.02,\\
 f_{\rho^0(770)}&=0.216\pm0.003,\quad f_{\rho^0(770)}^\perp=0.165\pm0.009,\\
f_{\bar{K}^*(892)}&=0.22\pm0.005,\quad f_{\bar{K}^*(892)}^\perp=0.185\pm0.010. \\
\end{split}
\end{equation}

As for the form factors, we use \cite{Cheng:2010yd,Cheng:2005nb}:
\begin{equation}
\begin{split}
F_0^{\bar{B}^0\rightarrow K}(0)&=0.35\pm0.04,\quad F_0^{\bar{B}^0\rightarrow f_0(500)}(m_K^2)=0.45\pm0.15,\quad A_0^{\bar{B}^0\rightarrow \rho^0(770)}(0)=0.303\pm0.029,\\
A_0^{\bar{B}^0\rightarrow \bar{K}^*(892)}(0)&=0.374\pm0.034, \quad F_0^{\bar{B}^0\rightarrow \pi}(0)=0.25\pm0.03.\\
\end{split}
\end{equation}

The values of Gegenbauer moments at $\mu=1 \mathrm{GeV}$ are taken from \cite{Cheng:2010yd,Cheng:2005nb},
\begin{equation}
\begin{split}
\alpha_1^\rho&=0,\quad \alpha_2^\rho=0.15\pm0.07, \quad \alpha_{1,\perp}^\rho=0,\quad \alpha_{2,\perp}^\rho=0.14\pm0.06, \\
\alpha_1^{K^*(892)}&=0.03\pm0.02,\quad \alpha_{1,\perp}^{K^*(892)}=0.04\pm0.03,\\
\alpha_2^{K^*(892)}&=0.11\pm0.09,\quad \alpha_{2,\perp}^{K^*(892)}=0.10\pm0.08,\\
B_{1,f_0(500)}^u&=-0.42\pm0.02,\quad B_{3,f_0(500)}^u=-0.58\pm0.19,\\
 B_{1,f_0(500)}^s&=-0.35\pm0.003,\quad B_{3,f_0(500)}^s=-0.43\pm1.26,\\
 B_{1,\bar{K}_0^*(700)}&=-0.92\pm0.11,\quad B_{3,\bar{K}_0^*(700)}=0.15\pm0.09.\\
\end{split}
\end{equation}
\end{appendix}

\acknowledgments
This work was supported by National Natural Science Foundation of China (Projects Nos. 11575023, 11775024, 11947001 and 11605150) and the Ningbo Natural Science Foundation (No. 2019A610067).

\end{document}